\documentclass[12pt,journal,compsoc]{IEEEtran}
\IEEEoverridecommandlockouts
\usepackage{float}
\usepackage{soul, color, acronym, multirow, amsmath, balance, amssymb, amsfonts}
\usepackage{cite}
\usepackage{graphicx}
\usepackage{textcomp}
\usepackage{xcolor}
\def\BibTeX{{\rm B\kern-.05em{\sc i\kern-.025em b}\kern-.08em
    T\kern-.1667em\lower.7ex\hbox{E}\kern-.125emX}}

\usepackage[justification=centering]{caption}
\usepackage{tikz, rotating, float, epsfig, subcaption, enumitem}
\usepackage{booktabs, textcomp, url}
\usepackage{pbox, listings, authblk, array, pifont}
\usepackage{optidef}
\usepackage{colortbl}

\usepackage{rotating, epsfig, subcaption, optidef}
\usepackage{listings, authblk, pifont}
\usepackage{algpseudocode}

\newcolumntype{C}[1]{>{\centering\arraybackslash}m{#1}}
\usepackage{soul, color, acronym, multirow,textcomp, amsmath, balance, amssymb, amsfonts}
\usepackage{cite}

\usepackage{geometry}
\usepackage{graphicx}
\usepackage[justification=centering]{caption}
\usepackage{tikz, rotating, float, epsfig, subcaption, enumitem}
\usepackage{booktabs, textcomp, url}
\usepackage{pbox, listings, authblk, array, pifont}
\usepackage{optidef}
\usepackage{colortbl}

\definecolor{color1}{rgb}{0.797,0.899,0.846}
\definecolor{color2}{rgb}{0.994,0.980,0.694}
\definecolor{color3}{rgb}{0.700,0.850,0.910}
\definecolor{color4}{rgb}{0.580,0.776,0.568}
\definecolor{color5}{rgb}{0.935,0.915,0.930}
\definecolor{color6}{rgb}{0.950,0.860,0.932}

\acrodef{PER}{Packet Erasure Rate}
\acrodef{PLR}{Packet Loss Rate}
\acrodef{PRR}{Packet Received Ratio}
\acrodef{BER}{Bit Error Rate}
\acrodef{RTT}{Round-Trip Time}
\acrodef{TCP}{Transmission Control Protocol}
\acrodef{UDP}{User Datagram Protocol}
\acrodef{AIMD}{Additive Increase Multiplicative Decrease}
\acrodef{PEP}{Performance Enhancing Proxy}
\acrodef{cwnd}{congestion window}
\acrodef{STP}{Satellite Transport Protocol}
\acrodef{BDP}{Bandwidth-Delay Product}
\acrodef{QoS}{Quality of Service}
\acrodef{LoS}{Line of Sight}
\acrodef{NLoS}{Non Line of Sight}
\acrodef{BLoS}{Beyond Line of Sight}
\acrodef{POMDP}{Partially Observable Markov Decision Process}
\acrodef{QoE}{Quality of Experience}
\acrodef{ACM}{Adaptive Coding and Modulation}
\acrodef{DAMA}{Demand Assignment Multiple Access}
\acrodef{VANET}{Vehicular Ad-Hoc Network}
\acrodef{RA}{Random Access}
\acrodef{CRA}{Contention Resolution ALOHA}
\acrodef{SA}{Slotted ALOHA}
\acrodef{RF}{Radio Frequency}
\acrodef{NG}{New Generation}
\acrodef{DSA}{Diversity Slotted ALOHA}
\acrodef{OBU}{On-Board Unit}
\acrodef{SVM}{Support Vector Machine}
\acrodef{LR}{Logistic Regression}
\acrodef{CRDSA}{Contention Resolution Diversity Slotted ALOHA}
\acrodef{SIC}{Successive Interference Cancellation}
\acrodef{ARQ}{Automatic Repeat reQuest}
\acrodef{SC-ARQ}{Selective-Coded ARQ}
\acrodef{V2V}{Vehicle to Vehicle}
\acrodef{SR-ARQ}{Selective-Repeat ARQ}
\acrodef{IRSA}{Irregular Repetition Slotted ALOHA}
\acrodef{CGC}{Complementary Ground Component}
\acrodef{GCS}{Ground Control Station}
\acrodef{RSU}{Road Side Unit}
\acrodef{ACK}{Acknowledgment}
\acrodef{NACK}{Negative Acknowledgment}
\acrodef{DVB-SH}{Digital Video Broadcasting - Satellite Services to Handhelds}
\acrodef{DVB-H}{Digital Video Broadcasting - Handheld}
\acrodef{DVB-RCS2}{Digital Video Broadcasting - Return Channel via Satellite, II generation}
\acrodef{SACK}{Selective Acknowledgment}
\acrodef{SNACK}{Selective Negative Acknowledgment}\acrodef{SNACK}{Selective Negative Acknowledgment}
\acrodef{MOS}{Mean Opinion Score}
\acrodef{SNIR}{Signal to Noise plus Interference Ratio}
\acrodef{SCPS-TP}{Space Communications Protocol Specifications - Transport Protocol}
\acrodef{CCSDS}{Consultative Committee for Space Data Systems}
\acrodef{ESA}{European Space Agency}
\acrodef{NASA}{National Aeronautics and Space Administration}
\acrodef{VAE}{Variational Autoencoder}
\acrodef{BSM}{Broadband Satellite Multimedia}
\acrodef{RLNC}{Random Linear Network Coding}
\acrodef{NC}{Network Coding}
\acrodef{FIFO}{First In, First Out}
\acrodef{FCFS}{First Come, First Served}
\acrodef{BLER}{Block Error Rate}
\acrodef{GEO}{geosynchronous}
\acrodef{LEO}{Low Earth Orbit}
\acrodef{FTP}{File Transfer Protocol}
\acrodef{CRC}{Cyclic Redundancy Check}
\acrodef{MAC}{Media Access Control}
\acrodef{PHY}{Physical layer}
\acrodef{HTTP}{Hypertext Transfer Protocol}
\acrodef{ISP}{Internet Service Provider}
\acrodef{MSS}{Maximum Segment Size}
\acrodef{BIC}{Binary Increase Congestion control}
\acrodef{AQM}{Active Queue Management}
\acrodef{XCP}{eXplicit Control Protocol}
\acrodef{CDF}{Empirical Cumulative Distributions}

\acrodef{ECN}{Explicit Congestion Notification}
\acrodef{CA}{Congestion Avoidance}
\acrodef{RED}{Random Early Detection}
\acrodef{TD}{Triple-Duplicate}
\acrodef{TO}{TimeOut}
\acrodef{IP}{Internet Protocol}
\acrodef{WMN}{Wireless Mesh Network}
\acrodef{ssthresh}{Slow-Start threshold}
\acrodef{MPE-IFEC}{Multi Protocol Encapsulation - Inter-burst Forward Error Correction}
\acrodef{FEC}{Forward Error Correction}
\acrodef{ML}{Machine Learning}
\acrodef{CRC}{Cyclic Redundancy Check}
\acrodef{P2P}{Peer-to-Peer}
\acrodef{FMT}{Fade Mitigation Technique}
\acrodef{SGD}{Smart Gateway Diversity}
\acrodef{NCC}{Network Control Centre}
\acrodef{ModCod}{Modulation and Coding}
\acrodef{FIFO}{First-In-First-Out}
\acrodef{WRR}{Weighted Round Robin}
\acrodef{WFQ}{Weighted Fair Queuing}
\acrodef{NS}{Network Simulator}
\acrodef{GSE}{Generic Stream Encapsulation}
\acrodef{PDF}{Probability Density Function}
\acrodef{CDF}{Cumulative Density Function}
\acrodef{CoV}{Coefficient of Variation}
\acrodef{MSC}{Message Sequence Chart}
\acrodef{ESA}{European Space Agency}
\acrodef{LIU}{Lebanese International University}
\acrodef{TUM}{Technical University of Munich}
\acrodef{MSCE-CS}{Master of Science in Communications Engineering - Communications Systems}
\acrodef{DLR}{German Aerospace Center}
\acrodef{NOS}{Network Operating System}
\acrodef{NFs}{Network Functionalities}
\acrodef{NC-SGD}{Network Coding for SGD}
\acrodef{ATSP}{Advanced Transport Satellite Protocol}
\acrodef{STP}{Satellite Transport Protocol}
\acrodef{WMN}{Wireless Mesh Network}
\acrodef{SNR}{Signal-to-Noise Ratio}
\acrodef{SINR}{Signal-to-Interference-plus-Noise Ratio}
\acrodef{GNNs}{Generative Neural Networks}
\acrodef{LMS}{Land Mobile Satellite}
\acrodef{LTE}{Long-Term Evolution}
\acrodef{M2M}{machine-to-machine}
\acrodef{IoT}{Internet of Things}
\acrodef{GAN}{Generative Adversarial Network}
\acrodef{IoT}{Internet of Things}
\acrodef{IoRT}{Internet of Remote Things}
\acrodef{IoST}{Internet of Space Things}
\acrodef{MIoT}{Multimedia Internet of Things}
\acrodef{RA}{Random Access}
\acrodef{UAV}{Unmanned Aerial Vehicle}
\acrodef{UTV}{Unmanned Terrestrial Vehicle}
\acrodef{UAS}{Unmanned Aerial System}
\acrodef{FANET}{Flying Ad-Hoc Network}
\acrodef{MANET}{Mobile Ad-Hoc Network}
\acrodef{VANET}{Vehicle Ad-Hoc Network}
\acrodef{C2}{Command and Control}
\acrodef{DTN}{Delay Tolerant Network}
\acrodef{COTS}{Commercial Off-the-Shelf}
\acrodef{IETF}{Internet Engineering Task Force}
\acrodef{CoAP}{Constrained Application Protocol}
\acrodef{MQTT}{Message Queuing Telemetry Transport}
\acrodef{URI}{Uniform Resource Identifier}
\acrodef{PUB/SUB}{Publish / Subscribe}
\acrodef{RCST}{Return Channel Satellite Terminal}
\acrodef{TDMA}{Time Division Multiple Access}
\acrodef{FDMA}{Frequency Division Multiple Access}
\acrodef{TCDMA}{Turbo Code Division Multiple Access}
\acrodef{PDMA}{Power Division Multiple Access}
\acrodef{WSN}{Wireless Sensor Network}
\acrodef{REST}{Representational State Transfer}
\acrodef{EDGE}{Enhanced Data rates for GSM Evolution}
\acrodef{UMTS}{Universal Mobile Telecommunications System}
\acrodef{LTE}{Long-Term Evolution}
\acrodef{E2E}{End-to-End}
\acrodef{3WHS}{Three-way Handshake}
\acrodef{SCADA}{Supervisory Control And Data Acquisition}
\acrodef{SOA}{Service-Oriented Architecture}
\acrodef{WebRTC}{Web Real-Time Communications}
\acrodef{fps}{frames per second}
\acrodef{SSIM}{Structural SIMilarity}
\acrodef{PSNR}{Peak Signal-to-Noise Ratio}
\acrodef{RPi}{Raspberry Pi}
\acrodef{NFV}{Network Function Virtualization}
\acrodef{OPEX}{Operating Expenditures}
\acrodef{CAPEX}{Capital Expenditures}
\acrodef{MIMO}{multiple-input and multiple-output}
\acrodef{SDN}{Software Defined Networking}
\acrodef{MTC}{Machine-type Communications}
\acrodef{mMTC}{Massive Machine-type Communications}
\acrodef{HTC}{Human-type Communications}
\acrodef{D2D}{device to device}
\acrodef{IIoT}{industrial IoT}
\acrodef{LPWAN}{Low-Power Wide-Area Network}
\acrodef{CPS}{Cyber-Physical System}
\acrodef{ICT}{Information and Communication Technologies}
\acrodef{SDR}{Software Defined Radio}
\acrodef{GPS}{Global Positioning System}
\acrodef{H2M}{human-to-machine}
\acrodef{MC}{Machine}
\acrodef{LA}{Local Area}
\acrodef{WA}{Wide Area}
\acrodef{HAP}{High Altitude Platform}
\acrodef{LAP}{Low Altitude Platform}
\acrodef{RAN}{Radio Access Network}
\acrodef{EIRP}{Equivalent Isotropically Radiated Power}
\acrodef{LT}{Logistics and Transportations}
\acrodef{BVLoS}{Beyond Visual Line of Sight}
\acrodef{BLoS}{Beyond Line of Sight}
\acrodef{VLoS}{Visual Line of Sight}
\acrodef{UTM}{Unmanned Aircraft System Traffic Management}
\acrodef{eNB}{Evolved Node B}
\acrodef{CDN}{Content Distribution Network}
\acrodef{ITU}{International Telecommunication Union}
\acrodef{SIN}{Space Information Network}
\acrodef{DTA}{Delay Tolerant Application}
\acrodef{DSA}{Delay Sensitive Application}
\acrodef{DTN}{Delay Tolerant Networking}
\acrodef{ISL}{Inter-Satellite Link}
\acrodef{TFRC}{TCP Friendly Rate Control}
\acrodef{RACH}{Random Access CHannel}
\acrodef{LPWAN}{Low-Power Wide Area Network}
\acrodef{VRT}{Variable Rate Technology}
\acrodef{HAP}{High Altitude Platform}
\acrodef{LALE}{Low Altitude Long Endurance}
\acrodef{LASE}{Low Altitude Short Endurance}
\acrodef{MALE}{Medium Altitude Long Endurance}
\acrodef{HALE}{High Altitude Long Endurance}
\acrodef{OTA}{Over The Air}
\acrodef{TLS}{Transport Layer Security}
\acrodef{SSL}{Secure Sockets Layer}
\acrodef{MEC}{Mobile Edge Computing}
\acrodef{GIS}{Geographic Information System}
\acrodef{GSM}{Global System for Mobile}
\acrodef{GPRS}{General Packet Radio Service}
\acrodef{PF}{Precision Farming}
\acrodef{SF}{Smart Farming}
\acrodef{SAR}{Syntethic Aperture Radar}
\acrodef{EO}{Earth Observation}
\acrodef{GNSS}{Global Navigation Satellite System}
\acrodef{EGNOS}{European Geostationary Navigation Overlay Service}
\acrodef{API}{Application Programming Interface}
\acrodef{RTP}{Real-time Transport Protocol}
\acrodef{RTCP}{Real-time Transport Control Protocol}
\acrodef{RR}{Receiver Report}
\acrodef{RB}{Round Robin}
\acrodef{WRR}{Weighted Round Robin}
\acrodef{ULFEC}{Upper Layer Forward Error Correction}
\acrodef{IC}{Interference Cancellation}
\acrodef{NIC}{Network Interface Card}
\acrodef{SCTP}{Stream Control Transmission Protocol}
\acrodef{ULPFEC}{Uneven Level Protection FEC}
\acrodef{DTMC}{Discrete Time Markov Chain}
\acrodef{GoP}{Group of Pictures}
\acrodef{NAT}{Network Address Translation}
\acrodef{NTN}{Non-Terrestrial Network}
\acrodef{MDP}{Markov Decision Process}
\acrodef{URLLC}{Ultra-Reliable and Low Latency Communications}
\acrodef{eMBB}{enhanced Mobile BroadBand}
\acrodef{VR}{Virtual Reality}
\acrodef{HTS}{High Throughput Satellite}
\acrodef{SN}{Sequence Number}
\acrodef{PAN}{Path-Aware Networking}
\acrodef{E2E}{End-to-End}
\acrodef{CSI}{Channel State Information}
\acrodef{PoC}{Proof of Concept}
\acrodef{AC}{Actor-Critic}
\acrodef{RL}{Reinforcement Learning}
\acrodef{UE}{User Equipment}
\acrodef{DQ}{Deep-Q}
\acrodef{NN}{Neural Network}
\acrodef{HIL}{Human In the Loop}
\acrodef{DLR}{Deep Reinforcement Learning}
\acrodef{POMDP}{Partially Observable Markov Decision Process }

\acrodef{GNN}{Generative Neural Network}
\acrodef{LOS}{Line-of-Sight}
\acrodef{NLOS}{Non-Line-of-Sight}
\acrodef{3GPP}{Third Generation Paternship Project}
\acrodef{DL}{Deep Learning}
\acrodef{AI}{Artificial Intelligence}
\acrodef{UAS}{Unmanned Aerial System}
\acrodef{HAP}{High Altitude Platform}
\acrodef{5G}{fifth generation}
\acrodef{6G}{sixth generation}
\acrodef{NR}{New Radio}
\acrodef{DU}{Distributed Unit}
\acrodef{CN}{Core Network}
\acrodef{CU}{Centralized Unit}
\acrodef{CFLOS}{Cloud Free Line-of-Sight}
\acrodef{QOS}{Quality of Service}

\acrodef{RL}{Reinforcement Learning}
\acrodef{UE}{User Equipment}
\acrodef{AC}{Actor-Critic}
\acrodef{MPR}{Multi-Path Routing}

\acrodefplural{eNB}[eNBs]{Evolved Nodes B}
\acrodefplural{NIC}[NICs]{Network Interface Cards}
\acrodefplural{RSU}[RSUs]{Road Side Units}
\acrodefplural{VANET}[VANETs]{Vehicular Ad-Hoc Networks}
\acrodefplural{OBU}[OBUs]{On-Board Units}
\acrodefplural{GNN}[GNNs]{Generative Neural Network}
\acrodefplural{UAV}[UAVs]{Unmanned Aerial Vehicles}
\acrodefplural{FANET}[FANETs]{Flying Ad-Hoc Networks}
\acrodefplural{MANET}[MANETs]{Mobile Ad-Hoc Networks}
\acrodefplural{VANET}[VANETs]{Vehicle Ad-Hoc Networks}
\acrodef{CDF}{Cumulative Distributions Function}

\acrodefplural{UAS}[UASs]{Unmanned Aerial Systems}
\acrodefplural{HAP}[HAPs]{High Altitude Platforms}
\acrodefplural{LALE}[LALEs]{Low Altitude Long Endurance}
\acrodefplural{LASE}[LASEs]{Low Altitude Short Endurance}
\acrodefplural{MALE}[MALEs]{Medium Altitude Long Endurance}
\acrodefplural{VAE}[VAEs]{Variational Autoencoders}
\acrodefplural{HALE}[HALEs]{High Altitude Long Endurance}
\acrodefplural{RCST}[RCSTs]{Return Channel Satellite Terminals}

\acrodefplural{GNN}[GNNs]{Generative Neural Networks}
\acrodef{CTGAN}{Conditional Tabular Generative Adversarial Network}
\acrodef{GAN}{Generative Adversarial Network}
\acrodef{TVAE}{Tabular Variational Autoencoder}

\newcommand{\pc}[1]{\iftrue \textcolor{blue}{\textbf{(PC: #1)}} \fi}
\newcommand{\ag}[1]{\iftrue \textcolor{magenta}{\textbf{(AG: #1)}} \fi}
\newcommand{\cg}[1]{\iftrue \textcolor{red}{(CG: #1)} \fi}
\newcommand{\ga}[1]{\iftrue \textcolor{green}{(GA: #1)} \fi}
\begin{document}
%
\title{
How Generative Models Improve LOS Estimation in 6G Non-Terrestrial Networks}

\author{Saira Bano, Achilles Machumilane, Pietro Cassarà, Alberto Gotta
\thanks{A. Machumilane and S.Bano are with the  Department of Information Engineering (DII) of the University of Pisa and the Institute of Information Science and Technologies (ISTI), CNR, Pisa - e-mails: {name.surname\}@isti.cnr.it}. A. Gotta and  P. Cassarà are with the Institute of Information Science and Technologies (ISTI), CNR, Pisa - e-mails: {name.surname}@isti.cnr.it}

}

\maketitle
\begin{abstract}
With the advent of 5G and the anticipated arrival of 6G, there has been a growing research interest in combining mobile networks with Non-Terrestrial Network platforms such as low earth orbit satellites and Geosynchronous Equatorial Orbit satellites to provide broader coverage for a wide range of applications. However, integrating these platforms is challenging because Line-Of-Sight (LOS) estimation is required for both  inter satellite and satellite-to-terrestrial segment links. Machine Learning (ML) techniques have shown promise in channel modeling and LOS estimation, but they require large datasets for model training, which can be difficult to obtain. In addition, network operators may be reluctant to disclose their network data due to privacy concerns. Therefore, alternative data collection techniques are needed. In this paper, a framework is proposed that uses generative models to generate synthetic data for LOS estimation in non-terrestrial 6G networks. Specifically, the authors show that generative models can be trained with a small available dataset to generate large datasets that can be used to train ML models for LOS estimation. Furthermore, since the generated synthetic data does not contain identifying information of the original dataset, it can be made publicly available without violating privacy. 
\end{abstract}

\begin{IEEEkeywords}
NTNs, Satellites, Channel Modeling, Generative Models.
\end{IEEEkeywords}
\section{Introduction}
 In recent years, the \ac{3GPP} has envisioned the integration of 5G and 6G mobile networks with \ac{NTN} technologies such as \ac{LEO} satellites, \acp{UAS}, and \acp{HAP} as a promising solution for providing ubiquitous coverage in inaccessible areas \cite{3gpp}.
 This integration will significantly improve network connectivity, accessibility and data rates and will also support a wide range of applications and services, including rescue missions, remote monitoring, and goods delivery \cite{3gpp}. The main communications challenge with this integration is modeling the \ac{LOS} availability of the link between the satellite and terrestrial segments since satellite communications require a clear \ac{LOS} that can be blocked by obstacles such as buildings and vegetation, resulting in signal blockage, diffraction, and reflection. The elevation angle of the satellite also affects the LOS, with lower angles less likely to result in a \ac{LOS} that yields to weak or no signal.
 Existing 3GPP and \ac{ITU} models define channel parameters based on elevation angle, frequency, and deployment scenarios \cite{9448842}. However, certain critical parameters, such as \ac{LOS} probability, have no temporal correlation and therefore do not account for satellite motion. Consequently, changes in \ac{LOS}/\ac{NLOS} states may be inaccurately represented, which could complicate the impact of radio mobility on 5G-based satellite networks. 
 Therefore, it is important to model these changes more accurately, considering the impact of satellite and user mobility. 
Statistical modeling is one potential solution for \ac{LOS} prediction and transmission of data only when conditions are favorable. However, this approach can be tedious and time-consuming. Recently, researchers have explored the use of \ac{ML} techniques for LOS estimation. Although these methods have shown promising results, they typically require large amounts of data to train the ML models, which can be difficult and expensive to obtain.
In addition, privacy concerns of network operators for sensitive data have made it difficult to obtain large data sets for \ac{ML} model training. To address this problem, generative models such as \ac{GNNs} have been proposed to generate synthetic datasets using small available real datasets. These networks have numerous applications in various fields, including communications, manufacturing, and healthcare. They are mainly used to create new images, music, text, and videos. The two most commonly used GNNs are \ac{GAN} and \ac{VAE}, which are used to generate synthetic data that mimics the quality and statistical distribution of real-world data. 
 This study aims to demonstrate the effectiveness of generating models in generating new satellite channel data for \ac{LOS} estimation. To accomplish this, the study utilizes the channel models provided by \ac{ITU} and creates datasets of \ac{LOS} and \ac{NLOS} traces for various satellite elevation angles that can account for satellite mobility. These datasets are then used to train the generative models, namely \ac{GAN} and \ac{VAE}. By using these models, this work shows that synthetic data can be generated that closely resembles the original data and retains the statistical distribution, thereby providing a solution to the limited availability of training data in \ac{ML} models for \ac{LOS} estimation. Furthermore the generated data would be free of legal, privacy, and security issues. People can use them for academic and research purposes and mitigate the difficulties of obtaining real-world data. We comprehensively evaluate, compare, and analyze the performance levels of the proposed generative models using various statistical parameters. The experimental results show that the proposed models are robust for LOS estimation using the generated dataset.

\section{Related Work}
This section provides various techniques proposed in the literature for \ac{LOS} estimation and synthetic data generation. For LOS estimation, the authors of \cite{bischel1996elevation} provide a theoretical model that calculates the likelihood of having LOS when there are no clouds. They showed that the LOS availability depends on the height of the ground station and the satellite elevation angle. A \ac{ML}-based method for NLOS identification is proposed in \cite{sun2022stacking} , which uses two \ac{ML} algorithms \ac{SVM} and \ac{LR} to predict the NLOS using data from the global navigation system. Authors in \cite{huang2020machine} use  time-varying angular information of a channel to train \ac{ML} algorithms for LOS identification in \ac{V2V} communication. The problem of these ML-based models is that they require a huge amount of measurement data that may not be available or too small to train a model.
Moreover, the measurement campaigns to obtain real channel data may be costly and time-consuming. Our work proposes a solution by showing that it is possible to use synthetic data instead of real channel data for LOS estimation. The synthetic data can be obtained by using generative models using a small amount of available real channel data or from publicly available synthetic data of the channel characteristics. 

Numerous data-driven methods for generating synthetic time series data are also studied in the literature. One example is the approach proposed by \cite{zhang2018generative}, which uses \ac{GAN} to generate energy consumption  data. \ac{GAN} uses a discriminator to indirectly train the generator network, enabling it to generate synthetic data. In \cite{geraci}, a general modeling approach is presented based on training a generative neural network with data. The proposed generative model includes two stages: Prediction of the link state (\ac{LOS}, \ac{NLOS}, or no available link) and subsequent use of a conditional \ac{VAE} to generate path losses, delays, and arrival and departure angles for all propagation paths given the previously predicted link state. In \cite{marey2022pl}, the authors used \ac{GAN} for path loss prediction for satellite images using the dataset from raytracing simulations.

\section{System Model}
\subsection{Channel Model}
The 3GPP has defined two main architectures for integrating cellular networks and NTN \cite{3gpp}: the transparent and regenerative architecture. In the transparent architecture, the satellite acts like a \ac{RF} repeater, transparently forwarding traffic between the UE and the \ac{NG}-\ac{RAN} of the mobile network. In the regenerative architecture, on the other hand, the satellite has gNB capabilities with onboard gNB-\ac{DU} and the gNB-\ac{CU} deployed in the ground segment of the mobile network, as shown in Figure \ref{fig1}. In this work, we use regenerative architecture. As explained earlier, the link between the satellite and the UE can be in LOS or NLOS. 

\begin{figure}[H]
\includegraphics[width=1\columnwidth,clip]{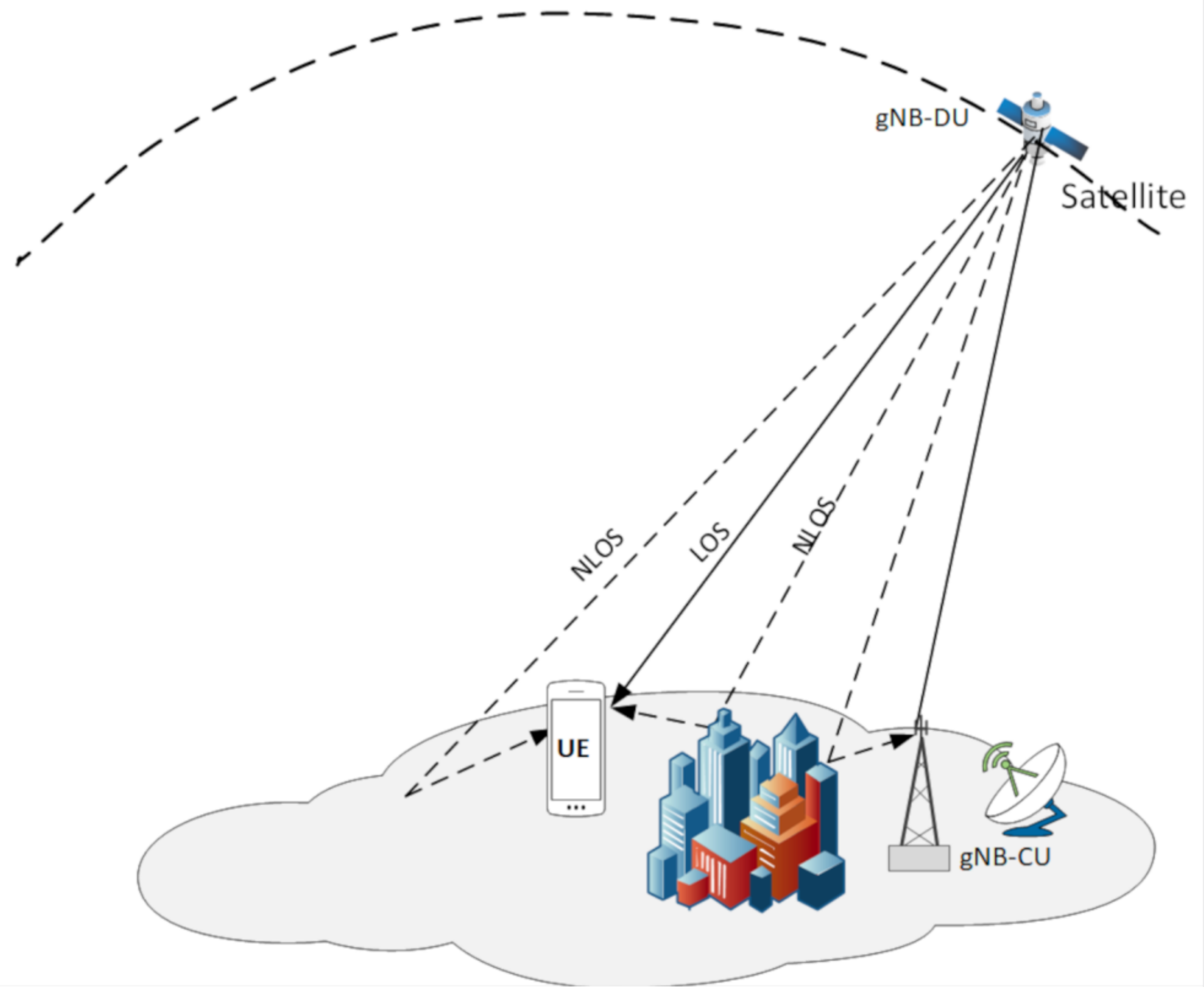}
\caption{Reference Scenario.\label{fig1}}
\end{figure} 

In this work, we investigate the generation of synthetic data that can be used to estimate the LOS probability for the link between the satellite and the UE. We use the channel model provided by the \ac{ITU} \cite{recommendation2017propagation} and modify it to a simplified Lutz model \cite{lutz1991land} , \cite{bischel1996elevation}. According to these models, a channel between a satellite and any land mobile terminal can be in either a good (G) or bad (B) state, as the signal power varies as a result of shadowing and multipath caused by signal obstructions and reflections from obstacles such as buildings, vegetation, and the ground. We assume that a channel is in a good state if there is a LOS, and in a bad state otherwise. The \ac{ITU} \cite{recommendation2017propagation} recommendations provide several statistical parameters that can be used to calculate the average duration of each of the two states in different environments, including the type of terrain (urban, suburban, and rural), elevation angles, and frequencies. These parameters were collected in a city in France and are shown in Table \ref{table1}. They include each state's mean, standard deviation, and minimum state lengths in meters.
For this study, we utilized the parameters for a dense urban environment at 2.2 GHz and computed the transition probabilities, which denote the likelihood of transitioning from one state to another given an initial state. Table \ref{table2} lists the transition probabilities we derived, and we modeled the state transitions using a two-state Markov process. To train our generative models, we generated LOS-NLOS traces from the statistical model. The generative models learned the latent space and distributions of the training data, allowing them to produce synthetic traces with similar statistical characteristics to the original dataset. Alternatively, without access to a channel model, the training data could be obtained from traces acquired through real-time transmissions using feedback mechanisms, as demonstrated in \cite{machumilane2022path}

\begin{table*}[h]
\centering
\caption{Satellite link parameters and transition probabilities for the dense urban environment in France at 2.2 GHz.\label{table1}}

\resizebox{1\textwidth}{!}{
\begin{tabular}{cccccc}
\toprule
Elevation	& $\mu_{G,B}$	& $\sigma_{G,B}$ &$dur_{minG,B}$&$P(B \to\, G)$ (g)	& $P(G \to\, B) $ (b)\\
\midrule

$20^\circ$ & 2.0042, 3.6890 & 1.2049, 0.9796 & 3.9889, 10.3114&0.00014310 &0.00047466
\\
$30^\circ$ & 2.7332, 2.7582 & 1.1030, 1.2210 & 7.3174, 5.7276&0.00024460 &0.00027570
\\
$45^\circ$ & 3.0639, 2.9108 & 1.6980, 1.2602 & 10.0, 6.0& 0.00020318 & 0.00007556
\\
$60^\circ$ & 2.8135, 2.0211 & 1.9595, 0.6568 & 10.0, 1.9126& 0.00105161 & 0.00010797
\\
$70^\circ$ & 4.2919, 2.1012 & 2.4703, 1.0341 & 118.3312, 4.8569& 0.00052923 & $2.76683\cdot10^-6$
\\
\bottomrule
\end{tabular}
}
\end{table*}









\subsection{Generative Neural Networks}

\subsubsection{Generative Adversarial Networks (GANs)}
In \cite{goodfellow2020generative} Ian Goodfellow introduces \ac{GAN}, a type of \ac{ML} algorithm that has gained popularity because of its ability to generate high-quality synthetic images, videos, and even sounds that closely resemble real-world data. GANs use an unsupervised learning approach to detect patterns in the input data and generate new samples with the same distribution as the original data. GANs train two neural networks, the generator network and the discriminator. The generator network generates the fake data by taking samples from a random distribution and converting them into data that resembles real data. In contrast, the discriminator network tries to distinguish between the real data and the fake data generated by the generator. The generator does not have direct access to real samples and learns only through interaction with the discriminator, which has access to synthetic and real samples. Training GANs is about finding the parameters of a discriminator that maximize its classification accuracy and finding the parameters of a generator that maximally confuse the discriminator. As the generator network improves, more realistic data is generated, making it increasingly difficult for the discriminator network to distinguish between real and fake data. This work uses GANs to generate LOS and NLOS traces for 6G-NTN channels.

\subsubsection{VAE - Variational Auto-Encoders}
Variational autoencoders are a powerful type of \ac{DL} algorithm that can be used for unsupervised learning and generative modeling. They are particularly useful for learning a compressed dataset representation that captures the data's underlying structure in a low-dimensional space. VAEs have many practical applications, such as image and speech recognition and natural language processing. VAEs are capable of generating new data that is similar to training data. 
VAEs use a two-part architecture consisting of an encoder and a decoder.
The encoder maps the input data to a distribution in latent space, while the decoder maps points in latent space back to the original data space. By varying the sampled points in the latent space, new data that is similar to, but not identical to, the training data can be generated. The VAE architecture includes an encoder that transforms input data into a Gaussian distribution over the latent space using convolutional or dense layers. The decoder takes samples from the latent space and maps them back to the original data space using similar layers. VAE aims to minimize the difference between input data and the decoder's output while maximizing the Gaussian distribution's parameter likelihood, which defines the latent space.

\subsubsection{Conditional Tabular GAN (CTGAN) and Tabular VAE (TVAE)}
This work uses Tabular Conditional GAN and Tabular VAE, both presented by Lei Xu et al. in \cite{xu2019modeling}, which are part of the Synthetic Data Vault (SDV) package in TensorFlow. These models were chosen because of their ability to process tabular data and allow training of a single model that can generate synthetic data for any available channel between UE and satellite. CTGAN and TVAE are capable of capturing the distribution of each column in the tabular data. In our case, each column contains LOS-NLOS traces for each channel or elevation angle. We consider three elevation angles: {$70^{\circ}$},{$60^{\circ}$},{$45^{\circ}$}. Knowing the distribution of each column, the models used in this work can generate synthetic data for each column simultaneously, based on the distribution of each column, which saves time compared to training a model for each channel. However, since the LOS probability changes with the elevation angle of the satellite, different channel models are needed for each angle. For \ac{LEO} satellites, this means that a different model is required for each elevation angle. Since satellites are visible from certain locations at certain angles, the terminal must switch to different satellites as they become visible. 
However, the proposed approach uses a single model that can generate synthetic data for all channels or satellite elevation angles that the UE connects to the satellite.

\section{Performance Evaluation}
In this part, we evaluate and compare the accuracy of the generative models in generating synthetic data that closely resemble real data. First, we train the models to generate synthetic data for the LOS and NLOS traces. Then we evaluate the trained models and compare the similarity between the generated data and the real data. We use Wasserstein Distance, Kolmogorov-Smirnov (KS) test, and Kullback-Leibler (KL) divergence as measures for comparison and evaluation. These measures allow us to determine how well the probability distribution of the synthetic data matches that of the real data.

\subsection{Model Training}
As explained previously, we trained our models using the training dataset consisting of the LOS-NLOS -traces obtained using the state transition probabilities in \ref{table1}. For this work, we used the probabilities at three elevation angles: $45^\circ$, $60^\circ$, $70^\circ$. We assume the satellite is visible from our reference UE at these three elevation angles. The trained models should produce synthetic data that can estimate or predict the presence of the LOS at these three angles. The training data set is a table with 100,000 rows of LOS and NLOS traces in three columns, with the traces of each angle in each column. We trained each model for 100 epochs with a batch size of 50 and a learning rate of $2e-4$.

\subsection{Performance Metrics}

We used the following performance metrics to evaluate the effectiveness of our geneative models for LOS estimation:
\begin{itemize}
    \item Wasserstein distance is a metric for measuring the distance between two probability distributions, that is, for measuring the similarity of the probability distribution of the synthetic data to that of the real data. The smaller the Wasserstein distance between the synthetic and real data, the more similar the probability distributions, which means that the generative model has produced synthetic data of high quality.
    \item The KS test is a statistical measure that gets the distance between two empirical \ac{CDF}, a popular non-parametric measure used in statistics. We use the KS test to measure how far the \ac{CDF} of the synthetic data is from the real data. It is usually presented as a complementary measure, i.e., a 1-difference in \ac{CDF}. Thus, the higher the value, the more similar the synthetic data is to the real data.
    \item The KL divergence measures how much two probability distributions differ from each other. The KL divergence between two probability distributions, P and Q, is calculated as the sum of the log difference between the probabilities of each value in P and Q multiplied by the probability of that value in P. The KL divergence is always non-negative and equal to zero if and only if the two distributions are identical. A low KL divergence indicates that the generative model has produced synthetic data of high quality that are similar to the real data.
\end{itemize}



\subsection{Results}

\begin{table}[h]
\centering
\caption{Distance between real and synthetic data for CTGAN.\label{table2}}
\resizebox{1\columnwidth}{!}{
\begin{tabular}{ccccccc}
\toprule
Metric&\multicolumn{2}{c}{$70^{\circ}$}&\multicolumn{2}{c}{$60^{\circ}$}&\multicolumn{2}{c}{$45^{\circ}$}\\

\midrule
&mean&variance&mean&variance&mean&variance \\

KS-test&0.9697&7.84e-08&0.9646&8.41e-08&0.9632&6.5025e-08\\
Wasserstein Distance&0.0591&4.096e-07&0.0700&1.21e-08&0.0743&7.48e-08\\
KL-Divergence&0.0199&7.85e-08&0.0101&1.42e-09&0.0034&6.87e-10\\

\bottomrule

\end{tabular}
}
\end{table}

\begin{table}[h]
\centering
\caption{Distance between real and synthetic data for TVAE.\label{table3}}
\resizebox{1\columnwidth}{!}{
\begin{tabular}{ccccccc}
\toprule

&\multicolumn{2}{c}{$70^{\circ}$}&\multicolumn{2}{c}{$60^{\circ}$}&\multicolumn{2}{c}{$45^{\circ}$} \\
\midrule
&mean&variance&mean&variance&mean&variance\\

KS-test&0.9651&9.6e-08&0.9763&8.1e-09&0.9570&1.156e-07\\
Wasserstein Distance&0.0591&4.096e-07&0.0701&1.21e-08&0.0743&7.84e-08\\
KL-Divergence&0.0247&8.969e-09&0.0039&4.242e-08&0.0052&7.904e-08\\
\bottomrule

\end{tabular}
}
\end{table}

\begin{figure*}

\includegraphics[width=.9\textwidth]{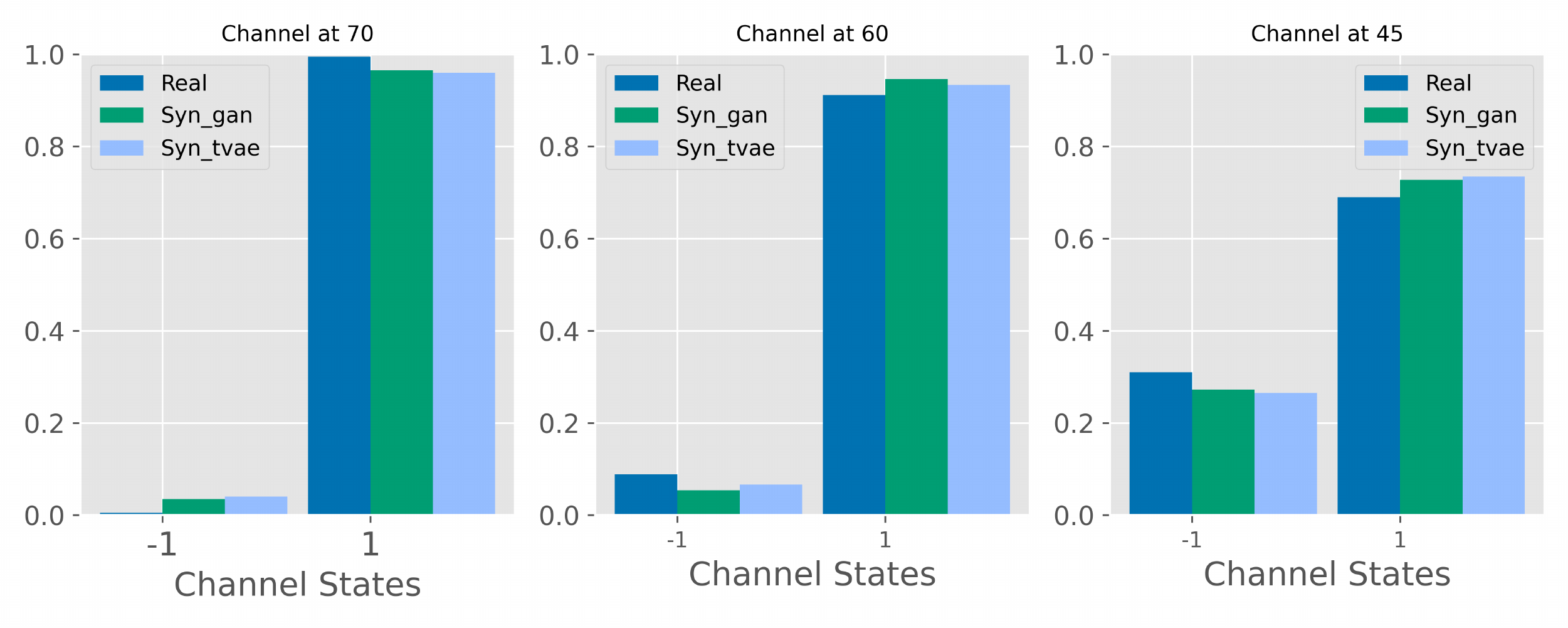}
\centering
\caption{Distribution of real vs synthetic data for CTGAN and TVAE at different satellite elevation angles.\label{fig1}}
\end{figure*} 

\subsubsection{Mean distance between real and synthetic data}
In Tables \ref{table2} and \ref{table3}, we show the mean and variance distance between synthetic and real data at different elevation angles using the selected evaluation metrics. To determine the mean distance, we first created a test dataset with 100,000 samples from LOS-NLOS, the same as the one created for the training dataset. Then we use the trained CTGAN and TVAE models to create synthetic datasets of the same size. We repeat this procedure 50 times and calculate the KS test, Wasserstein distance, and KL divergence distances between each synthetic data set and the real data, obtaining a set of fifty distances for each metric for each model. We then calculated the mean and variance of these distances, and the results are shown in Table \ref{table2} for CTGAN and Table \ref{table3} for TVAE. As can be seen from the tables, both the Wasserstein and KL divergences are very low for both models at all three elevation angles. Similarly, the values for the KS test are very high for both models, ranging from 0.9763 to 0.9570, indicating that our models produce high-quality synthetic data with distributions that are very close to the real data.
\subsubsection{Variance of the distance between real and synthetic data}
The variance of the distance between real and synthetic data shows whether the models can produce similar synthetic data given different specifications. The variance shows how stable and robust the models are in generating the synthetic data. The lower the variance, the higher the stability and robustness of the models. The results presented in \ref{table2} and \ref{table3} show low variances, as low as 1.42 x $10^-9$, which means that the distances between the different synthetic data sets generated in different instances vary little. This shows that our models are stable, robust, and reliable in generating LOS/NLOS estimates at different instances and different elevation angles.
\subsubsection{Distributions of real data and synthetic data}
In Figure \ref{fig1} we have shown the distribution of real and synthetic data for CTGAN and TVAE at different satellite elevation angles. In generating our dataset, LOS and NLOS were coded as 1 and -1, respectively. As expected, the LOS probability decreases with decreasing elevation angle and is highest at $70^{\circ}$ and lowest at $45^{\circ}$. The results show that this variation is similar for real and synthetic data, implying that our models can correctly estimate the LOS and NLOS probabilities at different elevation angles.
\subsubsection{Comparative Analysis}
\begin{figure}[H]
\includegraphics[width=1\columnwidth,clip]{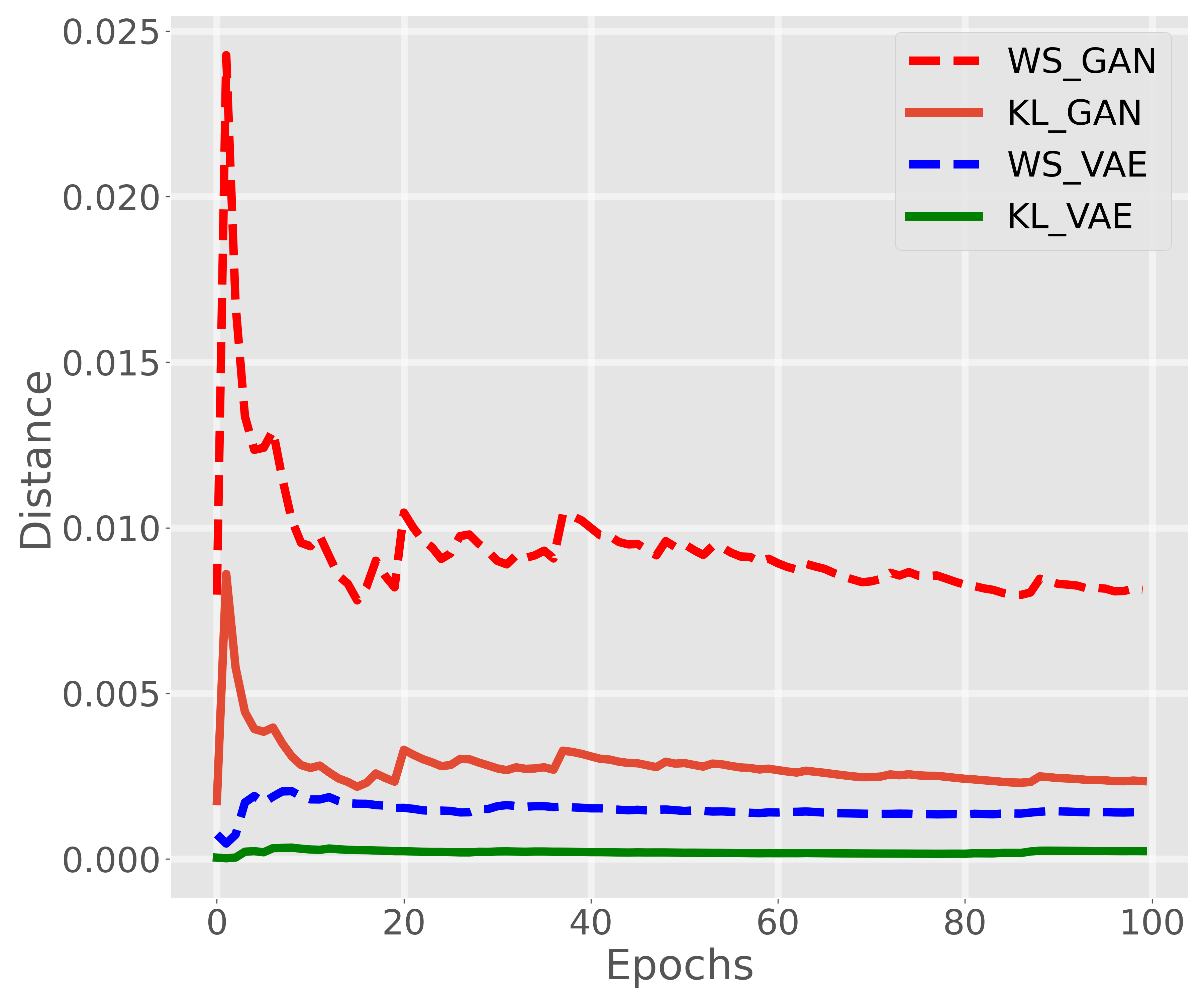}
\caption{Comparison of CTGAN and TVAE in terms of KL and Wasserstein distance against number of epochs for the channel at $70^{\circ}$\label{fig2}}
\end{figure}

The following briefly compares the generative models used in this paper. As mentioned earlier, both GANs and VAEs are \ac{DL} models used to generate synthetic data. However, they differ in the way they generate synthetic data. GANs use a generator that learns to generate synthetic data that is indistinguishable from real data.
In contrast, VAEs use a probabilistic encoder and decoder network to generate synthetic data by mapping the real data to a low-dimensional latent space and then mapping the latent space back to the original space. In general, GANs are known to be more difficult to train and may suffer from "mode collapse" compared to VAE, where the generator produces only a limited amount of synthetic data. The results in Figure \ref{fig2} show that VAEs exhibit faster stability with fewer runs over the training dataset, while GANs require longer training periods to achieve stability. The Figure shows the KL divergence and Wasserstein distance between the test dataset and the generated dataset for the elevation angle of $70^{\circ}$, indicating that the data distribution generated by the TVAEs is very similar to that of the real data set, even after only a few runs over the training dataset. Thus, the TVAEs perform better than the GAN for the LOS estimation for the given dataset. We also consider the training times of the CTGAN and TVAE. CTGAN requires more training time than TVAE. For example, in our simulations, CTGAN required 1.18 hours in the given training environment, while TVAE required 47 minutes. This shows that for the given problem of LOS estimation and with the given dataset, TVAE is the best option in terms of performance and training time.

\section{Conclusion}
\label{sec:conclusion}
In this study, a DL technique was used to generate synthetic data for LEO satellites operating in non-terrestrial 6G networks, considering both LOS and NLOS scenarios. Two generative network variants, CTGAN and TVAE, were used because they are well suited for tabular data. The simulation results showed that the generative models mimicked the real dataset very well and successfully estimated the LOS probabilities over multiple satellite channels. The statistical metrics used to measure the performance of the models showed that both models were comparable. However, TVAE outperformed CTGAN in terms of training time, while CTGAN exhibited some oscillations in training due to the min-max-adversarial effect.

\bibliographystyle{IEEEtran}
\balance
\bibliography{bibliography/bibliography}
\end{document}